# Heliophysics Event Knowledgebase for the *Solar Dynamics Observatory* and Beyond

(to appear in *Solar Physics*, SDO special issue. 2010)

N. Hurlburt, M. Cheung, C. Schrijver, L. Chang, S. Freeland, S. Green, C. Heck, A. Jaffey, A. Kobashi, D. Schiff, J.Serafin, R. Seguin, G. Slater, A. Somani and R. Timmons

*Lockheed Martin Advanced Technology Center, Palo Alto, CA, 94304, USA*

hurlburt@lmsal.com

http://www.lmsal.com/sungate

The immense volume of data generated by the suite of instruments on SDO requires new tools for efficient identifying and accessing data that is most relevant to research investigations. We have developed the Heliophysics Events Knowledgebase (HEK) to fill this need. The HEK system combines automated data mining using feature-detection methods and high-performance visualization systems for data markup. In addition, web services and clients are provided for searching the resulting metadata, reviewing results, and efficiently accessing the data. We review these components and present examples of their use with SDO data.

## 1. Introduction

The *Atmospheric Imaging Assembly* (AIA: Lemen et al. 2010) and the *Helioseismic and Magnetic Imager* (HMI: Schou et al. 2010) on the *Solar Dynamics Observatory* (SDO) represent a fundamental change in the approach to solar observations. Earlier solar missions were constrained by the technology of their time to follow two operational strategies that are both reflected in the instrumental precursors to AIA. Developers of the *Extreme ultraviolet Imaging Telescope* (EIT) project on SOHO (Delaboudinere *et al.,* 1995), for example, designed their investigations to observe the entire solar disk and adjacent corona at a regular sampling rate. This enables studies of global coronal dynamics and guaranteed complete coverage of all solar events, but sacrifices spatial and temporal resolution required to make detailed studies. As a complementary approach, the TRACE project (Handy *et al.,* 1999) designed their investigation to capture the dynamics of the corona in high temporal and spatial resolution. This limited the field-of-view of TRACE and forced observers to make careful plans to anticipate where and when interesting phenomena may occur.

The design of the AIA program removes the compromises of coverage *versus* resolution by observing the full disk at resolutions comparable to those of TRACE, with even better temporal coverage and improved spectral coverage. While this offers a dramatic improvement to the community, it shifts the operational problem from planning and coverage to one of data management. The data to be returned by AIA dwarfs that of TRACE and EIT. In one day, it delivers the equivalent of five years of TRACE data (2TB). To put the problem simply, the constraining factor for successful AIA science investigations is to efficiently find the data needed, and only the data needed for their purpose.

The Heliophysics Event Knowledgebase is designed to alleviate this problem. Its purpose is to catalog interesting solar events and features and to present them to the community in such a way that guides them to the most relevant data for their purposes. This is a problem of data markup that is arising in many scientific and other fields. Our approach shares many aspects with those from other fields. For example, the Monterrey Bay Aquarium Research Institute developed a similar system for annotating and cataloging video sequences of ocean fauna and activities (Cline *et al.,* 2007); various sports leagues have systems for cataloging clips of athletic events. The distinguishing factor for SDO is the large image format and complex event types. In the following sections we present the design and implementation of the HEK system and then present a short description of its use.

The design and implementation of the HEK and its interfaces allow expansion beyond its primary function as a searchable database that contains metadata on solar events and on the observation sets from which these are derived. It can, and does, track temporary data products that users may wish to have online for a limited period. It is also possible to include numerical data sets based on assimilated observations, or even such data sets and events within them that were generated as an observational data set would be. And, as a final example here, it can contain information on papers published based on certain data sets and events within them as associated metadata so that for any given event users can be pointed to such publications, or for a given publication users can be pointed to one or more data sets from one or more observatories.

## 1.1 HEK Design Goals

The Heliophysics Events Knowledgebase (HEK) is an integrated metadata system designed with the following goals in mind:

- Help researchers find data sets relevant to their topics of interest.
- Serve as an open forum where solar/heliospheric features and events can be reported and annotated.
- Facilitate discovery of statistical trends / relationships between different classes of features and events.
- Avoid overloading SDO data systems with attempts to download data sets too large to transmit over the internet.

To achieve these goals, HEK consists of registries to store metadata pertaining to observational sequences (Heliophysics Coverage Registry, or HCR), heliophysical events (Heliophysics Events Registry, or HER) and browse products such as movies. Interfaces for communications and querying between the different registries are also provided in terms of web services.

### *1.1.1 Heliophysical Events*

While many think of events as physical processes, it is more appropriate for our purposes to use a more empirical definition. This enables our system to include entries that come from multiple sources and methods, while avoiding ambiguities. Events can be grouped into two general categories: those that are directly found in data, and those that are inferred by other means. In the former case, any report of an event must specify an event class and the method, data and associated parameters used to identify the event as well as contributed the event. For example: “I” found a “flare” in “AIA 195 data” using the SolarSoft routine “ssw_flare_locator” with “default parameters”. The extensible list of possible event classes, shown in Table 1, is maintained as a controlled vocabulary by the HEK team. For each event class, a unique set of required and optional attributes is also defined by the HEK team. All events are required to have a duration and bounding box that contains the event in space and time. Some assumptions are made about locations dependent upon the event class. For instance, solar events are assumed to be on the solar surface unless explicitly stated otherwise. Similarly, heliospheric events seen from a single perspective, such as CMEs observed by SOHO/LASCO are taken to be on the plane of the sky containing the solar center.

The second category of events, namely those inferred by other means, can also be captured within the HEK. In this case, the required information is the inference method, the metadata that it operated upon, and the parameters it used for making the inference. For instance, Active Regions (ARs) are reported and assigned numbers daily by NOAA. Using our empirical definitions, each AR observation is an event bounded within a 24-hour time interval. An obvious inference is that all NOAA active regions with the same Active Region Number are the same active region. In this case, the data source is the Heliophysics Events Registry and the method is a query to the HER for Active Regions with a given Active Region number. Similar higher-level events can be constructed to the point where they can be matched against the “physical” counterpart and these can either by registered back into the HER or derived on the fly.

We have extended the VOEvent schema developed by the IVOA (White *et al.,* 2006) to encapsulate all metadata describing HEK events. This provides a convenient means for exchanging event descriptions between various providers and can be used as a means for distributed processing of events (Hurlburt *et al.* 2008). SolarSoft (Freeland and Bentley, 2000) routines have been developed which support HEK events .

| **Event Class** | **Description** |
|---|---|
| Active Region | Solar Active Region |
| Coronal Mass Ejection | Ejection of material from the solar corona |
| Coronal Dimming | A large-scale reduction in EUV emission |
| Coronal Jet | A jet-like object observed in the low corona |
| Coronal Wave | EIT or Morton waves spanning a large fraction of the solar disk |
| Emerging Flux | Regions of new magnetic flux in the solar photosphere |
| Filament | Solar Filament or Prominence |
| Filament Eruption | A sudden launching of a filament into the corona |
| Filament Activation | A sudden change in a filament without launching |
| Flare | Solar Flare |
| Loop | Magnetic loops typically traced out using coronal imagery |
| Oscillation | A region with oscillating coronal field lines |
| Sigmoid | S-shaped regions seen in soft X rays; indicator for flares |
| Spray Surge | Sudden or sustained intrusion of chromospheric material well into the corona |
| Sunspot | Sunspots on the solar disk |
| Plage | Bright areas associated with active regions |
| Other | Something that could not be classified – good candidate for further research |
| Nothing Reported | Used to indicate that the particular data were examined, but had nothing noteworthy to the observer. |

Table 1. The current list of events classes supported by the HEK. This is a controlled list that is expandable as new events classes are defined.

The concept of an “event” can easily be expanded. For instance, the VOEvent schema supports descriptions of observatory configurations at specified times as well. We have used these extended properties to capture metadata associated with instrument operations and incorporated them into a database that describes the type and purpose of data collected. These “Coverage Events” include information on who collected the data, what wavelengths were observed, and in what regions and time intervals the was data collected. Additional event classes are being defined to support the execution of models and simulations of heliophysical phenomena. In each case, the VOEvent standard permits easy exchange and transport of event descriptions. We have developed service and client software that permit these exchanges.

These descriptions can also be annotated after the fact, resulting in something akin to social-networking site for heliophysical data. It can be integrated with scientific publications by explicit links or references, such as those provided the Astrophysical Data System (http://ads.harvard.edu). Alternatively, they can be referenced implicitly using standard descriptions of events supported by publishers, such as the SOL convention that has been adopted by journals supporting solar physics publications.

### 1.2 Example usage

The research community interacts with the HEK by several methods. For example, iSolSearch (Figure 1) is a web application that presents a unified experience with the services provided by the HEK and SolarSoft. It permits web users to search for different event classes with a variety of constraints. Here, a set of common event types is selected, including flares CMEs and Active Regions. Events are displayed as icons on a solar disk or Carrington grid and as a list. There are a variety of options for saving these list for future use. When an individual event is selected in iSolSearch, a summary panel of the event opens with links to more detailed information including movies and images that help identify events worthy of further study. In addition, iSolSearch finds and displays observational sequences available in the vicinity of the selected event. Users follow these links to download the data containing the events. These links lead directly to the appropriate data sources or, when appropriate, pass the request onto the Virtual Solar Observatory.

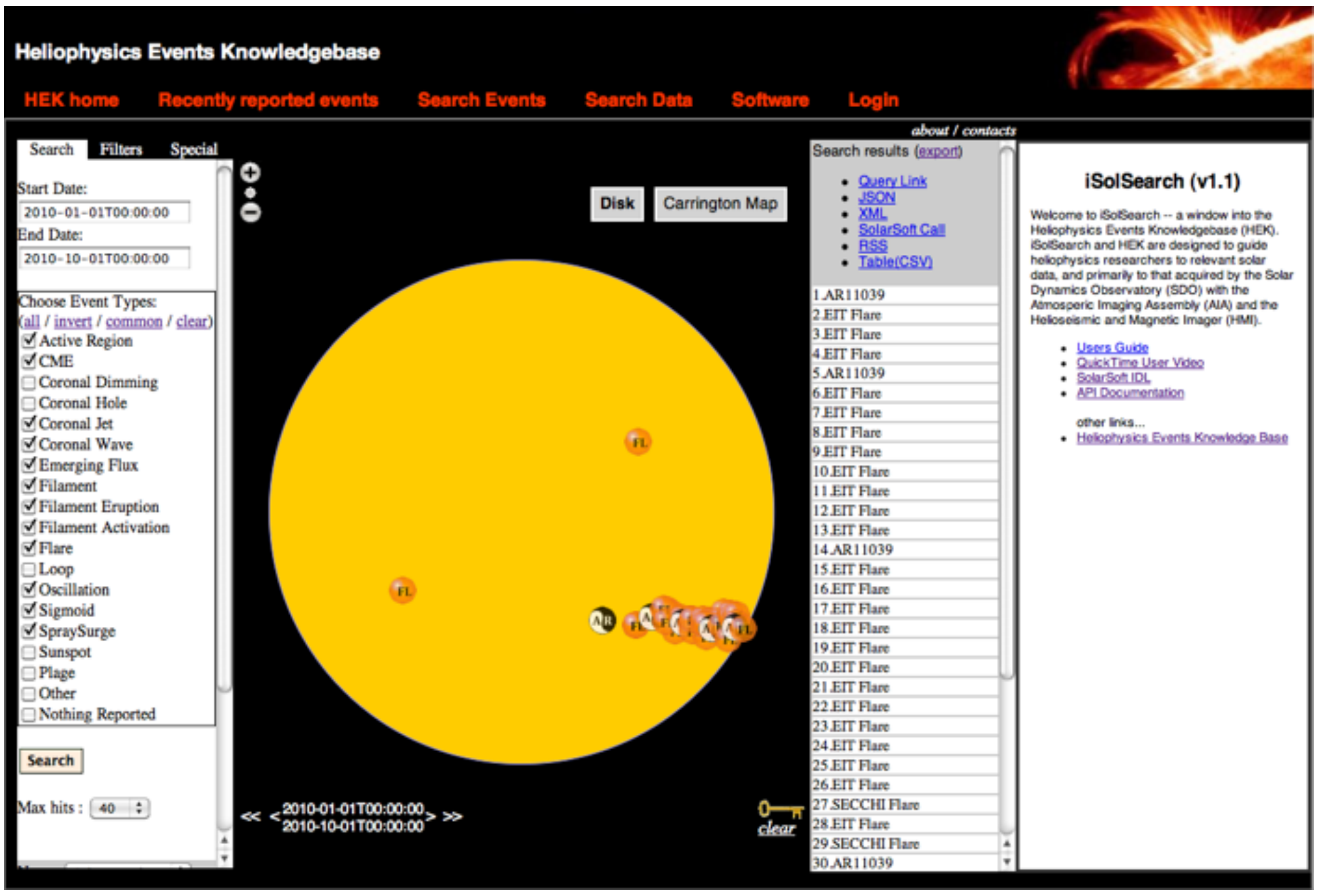

Figure 1. The iSolSearch client, http://lmsal.com/isolsearch, interacts with various HEK registries to present a unified view of events and available data.

An alternative approach to using the HEK is through SolarSoft. This snippet of IDL code uses the HEK-provided routines to find flares within a specified time interval:

```
IDL> t0='1-jan-2000' & t1='1-jan-2001'
IDL> query = ssw_her_make_query(t0,t1,/flare)
IDL> events = ssw_her_query(query)
```

Further analysis of these events can be fed into a search for data:

```
IDL> t0=events(0).starttime & t1=events(0).endtime
IDL> dataset = ssw_hcr_query(t0,t1)
```

## 2. HEK Implementation

We have developed a system for extracting events and features from SDO data and for presenting them to users in a convenient fashion. Here we review the system design and operation. The overall design is displayed in Figure 2, with major functional areas color coded for easy identification. These three areas encompass Mission Assets (green), primarily focused on missions such as TRACE and *Hinode* that have an extensive planning component; the Joint Science Operations Center (JSOC) Assets (gray), which are the data-oriented components involved in SDO/HMI and AIA operations; and Public Assets (red), which are primarily involved in presenting the outcomes of the first two to the general user community. Here we discuss each of these in more detail.

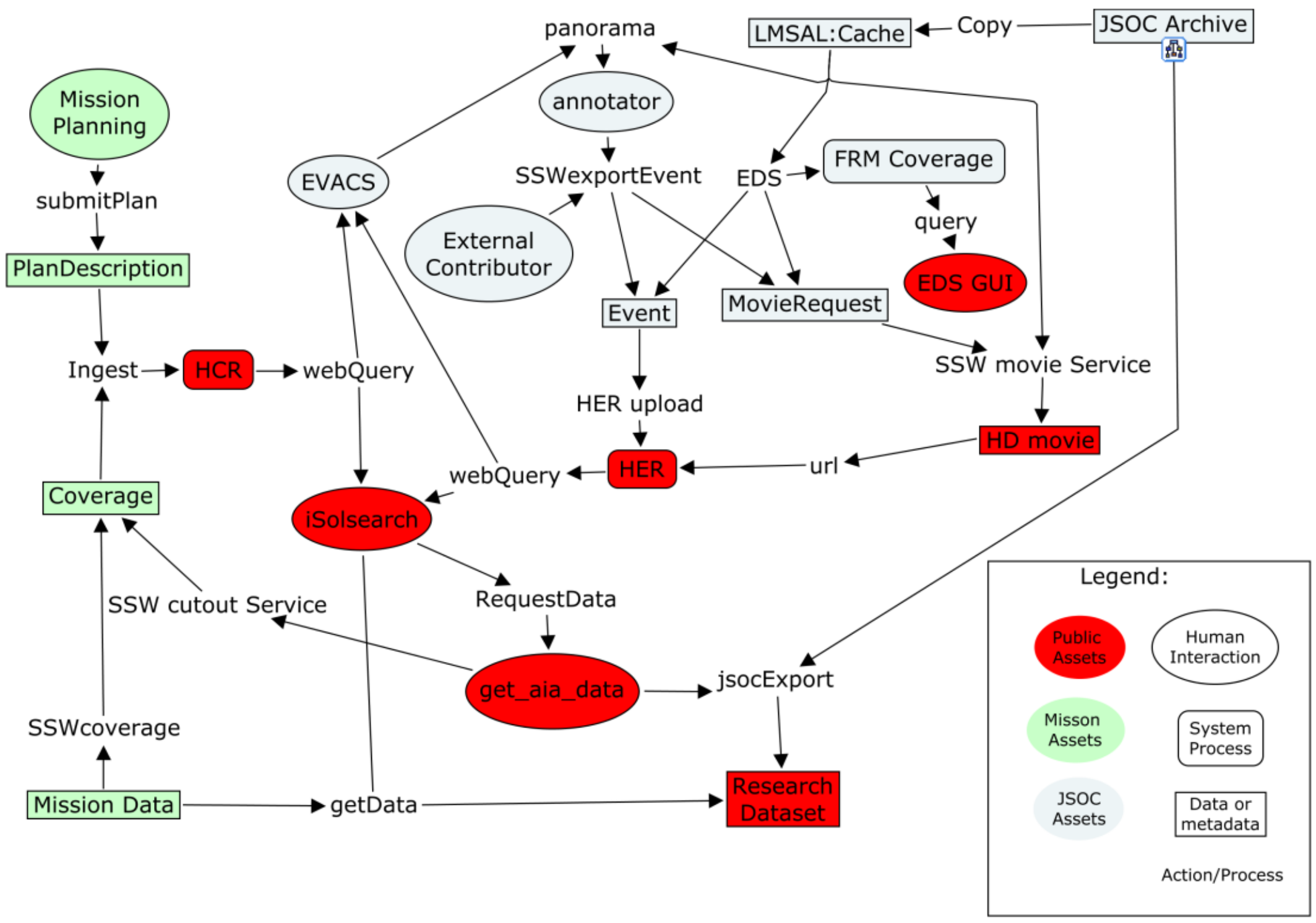

Figure 2. HEK Components include the Event and Coverage Registries (HER, HCR), Inspection and Analysis Tools, Event Identification System and Movie Processing. Event services enable web clients to interact with the HEK and use the results to request data from the JSOC at Stanford.

## 2.1 Mission (Coverage) Assets

### *2.1.1 Capturing Planning Metadata*

Metadata associated with TRACE and *Hinode* images and observations consist of detailed image catalogs which describe individual images and higher-level descriptions of image sets, including pre-observation notes from the planner or observer, and post-observation annotations by the TRACE, SOT, and XRT science teams. Metadata associated with the observation planner is captured at the time the daily observing plan is committed. A summary of the plan is extracted by automated tools which then registers these plan descriptions into the HCR. After the plans are executed and the resulting data are cataloged, the results are compared and cross-referenced to the plan.

## 2.2 JSOC Assets

Calibrated SDO image data are accessed from the Joint Science Operations Center Science Data Processing (JSOC/SDP) system at Stanford over a private link to a 100 TB cache at the LMSAL. This cache feeds into three components within the LMSAL facility: the Event Detection System, which operates on a dedicated computer cluster; the AIA Visualization Center, containing a high-performance display system; and an image-extraction and browse-product generation system.

### *2.2.1 Event Detection System*

The Event Detection System (EDS) autonomously orchestrates a variety of feature- and event-detection modules in order to populate the HEK with events from SDO data. The EDS continually acquires incoming data, provides the appropriate data to each module, receives results from modules through a standard API, and uploads the results to the HER. The outputs of the EDS are event descriptions in the VOEvent format (White *et al.,* 2006), including at a minimum the type, time, and location of the detection. Some modules produce ancillary data products and the EDS creates links from the HER to them, and the EDS also sends information to the movie-creation service to create summary clips for its events.

The Smithsonian Astrophysical Observatory-led Feature Finding Team (FFT) is the main source of modules for SDO (Martens *et al.,* 2010). However, the interfaces needed for modules to run in the EDS are published openly, and contributions from the wider solar-physics community are encouraged. Modules may run in a standalone mode or can be “triggered”, meaning they subscribe to certain types of VOEvents from other modules in the EDS and run only when and where those types of events are found. This mode is useful for feature recognition methods that are too computationally expensive to run on the entirety of SDO data.

The EDS consists of a single head node and a number of worker nodes connected via JXTA, a Java-based peer-to-peer networking technology (http://jxta.dev.java.net). At system initialization, the head node assigns modules to nodes. During operation, the head node monitors a cache and distributes announcements of the appropriate wavelengths/types of new data to the modules. Module results are sent back to the head node, which uploads them to the HER. The head node also tracks resource-usage statistics and can reassign modules to other nodes if there is a failure or an overload of one node's computing power. In addition, the head node can receive commands to update a module version or add in new modules during operations without restarting the rest of the system. The head node also maintains a Feature Recognition Method Coverage Registry that records when EDS modules have been run, what parameters and code versions were used by the modules, and which ones were triggered by others at what time. Besides internal record keeping, this registry permits identification of "All Clear" periods which could form the basis for space-weather prediction (using the HER alone, it would not be clear if an absence of events during a specified period was due to actual solar inactivity, or problems that prevented the modules from running during that time).

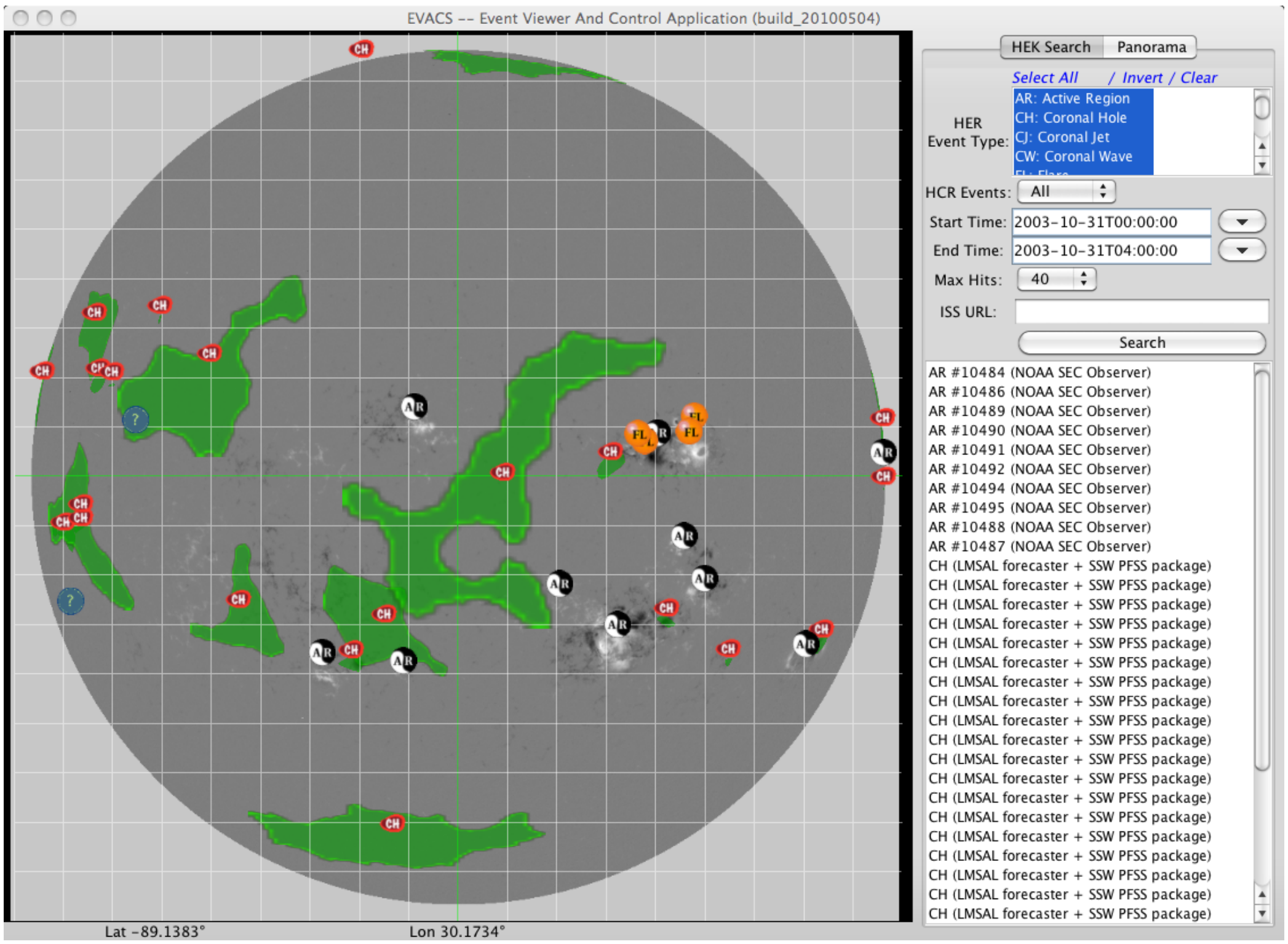

Figure 3. The Event Viewer and Control System control panel.

Some modules (*e.g.* those detecting flares, emerging flux, and coronal dimmings) produce data that is useful for immediate action. These modules can produce "open" events soon after first detection, denoting the beginning of some phenomena, which form the basis for space weather alerts or trigger immediate execution of other modules. When the event ends, the modules submit a "closed event" that updates the original in the HEK. These modules run on "quicklook" SDO data that arrive less than an hour after observation, while others may run on the final version that comes approximately a day later. The EDS stores this distinction and other information on data provenance in its VOEvents.

Currently there are two instances of the EDS, one at LMSAL and one at the Smithsonian Astrophysical Observatory (SAO); the latter includes some modules that produce auxiliary science products that is archived separately from the HEK. Although the EDS was built primarily to deal with the large data volume from SDO, it is applicable to other solar data sets as well; both H$\alpha$ observations and SOHO/LASCO data are planned to be used by FFT modules. We are considering extending it to scientific data processing applications outside of solar physics as well.

### *2.2.2 AIA Visualization Center*

In parallel with the automated EDS, all AIA data are inspected by the AIA science team. A dedicated inspection facility, with a large high-resolution display and viewing tools, is used to review and annotate the automated detections from EDS and to add events that have no reliable means of automated detection. The results are passed along in the same format as those from the EDS. Similar, but less intensive, data inspection can be made by the science team directly from their workstations, or by external team members. The inspection and browsing system consists of multiple tools that generate images, movies, and metadata which are then uploaded and stored into the HER and the HCR.

The Event Viewer And Control System (EVACS) tool displayed in Figure 3 uses metadata stored in the HER and HCR to display time and channel information for a given observation. EVACS controls a high-performance image-viewing tool called Panorama for use in reviewing the data and an annotation tool for reporting events. Panorama displays multiple channels of time-series of FITS or other image data. It can overlay them, pan and zoom spatially, and select regions in space and in time on a HiPerSpace data wall (Defanti *et al.,* 2009) such as that displayed in Figure 4. Once a desired sequence and event is identified, the user launches the Annotator tool from within Panorama or EVACS. Panorama takes full advantage of modern hardware-accelerated graphics including the CGLX library (Doerr and Kuester, 2009) for multi-tiled display and the performance available from today's graphics chipsets to achieve high-speed visualization of very large image sets.

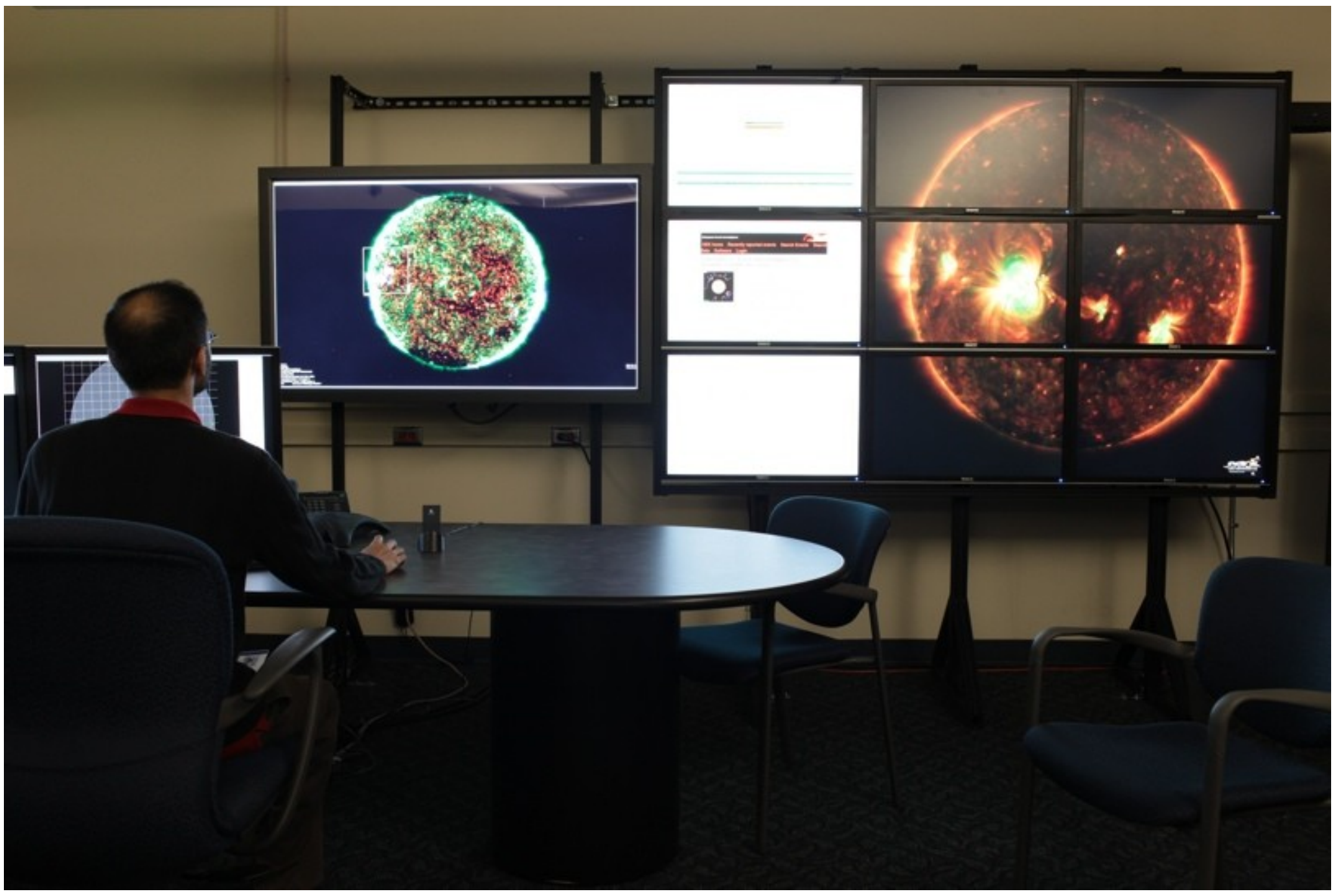

Figure 4. HiPerSpace data wall controlled by EVACS/Panorama.

The Annotator tool has an extensive user interface to assist AIA data analysts in describing observed events. Once all appropriate values are supplied, the Annotator generates VOEvent XML content and registers it in the HER. If the data being annotated is associated with a coverage event, appropriate linkage and citation between the events in the HER and HCR are created and preserved so users can browse in both directions from source observations to the derived data products and identified features.

### *2.2.3 Browse Products*

Movie processing takes the inputs from EDS and inspection tools to generate movie clips in a variety of formats and resolutions. The particular combination of wavelength, cadence, and resolution are dependent both on the method that found the data and the type of feature reported. The resulting movies and ancillary data are web accessible with references in the HER.

## 2.3 Public Assets

### *2.3.1 The Heliophysics Event and Coverage Registries*

A set of registries store the resulting outputs of all the analyses described above and form the core of publicly-accessible components of the HEK system. The Heliophysics Event Registry accepts the VOEvent files and registers their contents into a PostGIS database. The Heliophysics Coverage Registry contains descriptions of the instrument coverage, that is the "where," "when," and "how" of a particular observation. A variety of coordinate systems are supported for contributions based upon the World Coordinate System (Thompson, 2006).

### *2.3.2 HEK Services*

The HEK is presented through a web server in a variety of forms. It is closely tied to the Stanford website (http://jsoc.stanford.edu) and supports an interface for passing the results of event searches and movie browsing directly to the data request services developed at Stanford. Besides the standard web interface - such as presenting users with movies, pre-computed data products and static forms for interrogating the registries - the web server supports a variety of services that can be used by external teams to develop web applications.

### *2.3.3 HEK Clients*

Aside from iSolSearch and SolarSoft clients that have been discussed earlier, clients from other developers are encouraged. Two that are currently under development are those of the Helioviewer.org and jHelioviewer projects (http://helioviewer.org) and the Virtual Solar Observatory (Hill *et al.,* 2009). In addition, both EVACS and Panorama are distributed through SolarSoft in the Panorama package.

| Service Area | Service | Description | Result |
|---|---|---|---|
| **HER** | **search** | **Event search** | **Event list in a variety of formats** |
| | **view-event** | **Event display** | **Document describing event** |
| | **export-event** | **Reports back event details** | **VOEvent file or other format** |
| | **View attributes** | **Returns all values of attributes** | **list** |
| | **Home** | **HER home page** | **Displays recent events submitted by humans** |
| **HEK** | **Heks** | **Authenticated Services** | **https://www.lmsal.com/hek/heks** |
| | **Login** | **Login to HEK** | **authorization** |
| | **Comment** | **Comment on event** | **Records comment to existing event entry** |
| | **Rate** | **Rate event** | **Records rating of existing entry** |
| | **Add reference** | **Create link to event** | **Adds a URL link to existing entry** |
| **HCR** | **Search** | **Instrument coverage search** | **Coverage list** |
| | **Display** | **Details** | **Document describing instrument coverage** |
| | **Ingest** | **Accepts coverage descriptions** | **Registers instrument coverage in HCR** |
| **SolarSoft** | **SSWexportEvent** | **Create and upload event to HER** | **VOEvent file** |
| | **SSWcoverage** | **Create and upload to HCR** | **VOEvent file** |
| | **SSWcutout** | **Create cropped image sets** | **Web summaries** |
| | **General** | **Time and coordinate services** | **Converted units** |

Table 2. A summary of HEK web services. Complete documentation is available on the LMSAL SDO documentation site (http://www.lmsal.com/sdodocs).

### *2.3.4 Data request management*

One of the prime motivations for developing the HEK is to assist users in requesting subsets of the full image stream, rather than requesting large chunks of data that then requires winnowing on their part. This reduces both the network load and user overload. The last step in the use of the HEK is to get the data. This requires finding subsets of associated data with HER events and, when needed, ordering data based on field of view, time, resolution, and wavelength. A means to share and track these datasets is also useful to prevent repeated regeneration of popular datasets.

The HEK guides researchers to the most useful times and locations and will *suggest* reasonable fields of view and durations. With *guidance* from the HEK, researchers can access data by specifying cutouts, wavelengths, and sampling rates in space and time. These research data products use the best flat fields and calibration data available at the time that they are created. They are tagged in the HCR. Subsequent researchers can be directed to existing, local copies of research products through the HEK.

# 3. Conclusions

We have developed the Heliophysics Event Knowledgebase to address the immediate needs of the *Solar Dynamics Observatory*. However, the underlying motivation of devising a means for coping with petabyte datasets resonates with many other missions and projects throughout modern science. Our system design and implementation were intended to grow into a wider effort, as suggested by the use of "Heliophysics" in its title, when SDO is primarily a solar mission. We hope our description here will encourage and benefit related efforts throughout the heliosphere and beyond. For more information visit http://lmsal.com/sungate.

Acknowledgements

We thank our partners at all of the institutions that have contributed to our efforts, most notably those at the Smithsonian Astrophysical Observatory and at the Royal Observatory of Belgium. This work was supported by NASA through grants NAS5-38099, NNM07AA01C, NNG04EA00C and Lockheed Martin Internal Research Funds.